\documentclass[twocolumn,aps,superscriptaddress,showpacs]{revtex4}

\usepackage{amssymb}
\usepackage{amsmath}
\usepackage{graphicx}
\usepackage[normalem]{ulem}
\usepackage[dvips]{color}
\usepackage{multirow}
\usepackage{appendix}

\begin{document}

\title{Constraining simultaneously nuclear symmetry energy and neutron-proton effective mass splitting with nucleus giant resonances from a dynamical approach}
\author{Hai-Yun Kong}
\affiliation{Shanghai Institute of Applied Physics, Chinese Academy
of Sciences, Shanghai 201800, China}
\affiliation{University of Chinese Academy of Sciences, Beijing 100049, China}
\author{Jun Xu\footnote{corresponding author: xujun@sinap.ac.cn}}
\affiliation{Shanghai Institute of Applied Physics, Chinese Academy
of Sciences, Shanghai 201800, China}
\author{Lie-Wen Chen}
\affiliation{Department of Physics and Astronomy and Shanghai Key
Laboratory for Particle Physics and Cosmology, Shanghai Jiao Tong
University, Shanghai 200240, China} \affiliation{Center of
Theoretical Nuclear Physics, National Laboratory of Heavy Ion
Accelerator, Lanzhou 730000, China}
\author{Bao-An Li}
\affiliation{Department of Physics and
Astronomy, Texas A$\&$M University-Commerce, Commerce, TX
75429-3011, USA} \affiliation{Department of Applied Physics, Xi'an
Jiao Tong University, Xi'an 710049, China}
\author{Yu-Gang Ma}
\affiliation{Shanghai Institute of Applied Physics, Chinese Academy
of Sciences, Shanghai 201800, China}
\affiliation{Shanghai Tech University, Shanghai 200031, China}

\date{\today}

\begin{abstract}
With a newly improved isospin- and momentum-dependent interaction
and an isospin-dependent Boltzmann-Uehling-Uhlenbeck transport
model, we have investigated the effects of the slope parameter $L$
of the nuclear symmetry energy and the isospin splitting of the
nucleon effective mass $m_{n-p}^*=(m_n^*-m_p^*)/m$ on the centroid
energy of the isovector giant dipole resonance and the electric
dipole polarizability in $^{208}$Pb. With the isoscalar nucleon
effective mass $m_s^*=0.7m$ constrained by the empirical optical
potential, we obtain a constraint of $L=64.29\pm11.84 (\rm MeV)$ and
$m_{n-p}^*= (-0.019 \pm 0.090)\delta$, with $\delta$ being the
isospin asymmetry of nuclear medium. With the isoscalar nucleon
effective mass $m_s^*=0.84m$ extracted from the excitation energy of
the isoscalar giant quadruple resonance in $^{208}$Pb, we obtain a
constraint of $L=53.85\pm10.29 (\rm MeV)$ and $m_{n-p}^*= (0.216 \pm
0.114)\delta$.

\end{abstract}


\maketitle

\section{Introduction}
\label{introduction}

One of the main tasks of nuclear physics is to understand the
in-medium nuclear interactions and the equation of state (EoS) of
nuclear matter. The uncertainties of the isospin-dependent part of
the EoS, i.e., the nuclear symmetry energy ($E_{sym}$), has hampered
our accurate understanding of nuclear matter properties, while it
has important ramifications in heavy-ion reactions, astrophysics,
and nuclear
structures~\cite{ireview98,ibook01,dan02,Bar05,Ste05,Lat07,Li08,EPJA}.
Thanks to the great efforts made by nuclear physicists in the past
decade, a more stringent constraint on $E_{sym}$ at subsaturation
densities has been obtained from various analysis, with the slope
parameter of the nuclear symmetry energy so far constrained within
$L=60 \pm 20$ MeV~\cite{Che12,BAL13,Oer16}, although further
verifications are still needed. On the other hand, the in-medium
isospin splitting of the nucleon effective mass
$m_{n-p}^{\ast}=(m_n^*-m_p^*)/m$ has become a hot topic recently.
Compared to the bare nucleon mass in free space, the in-medium
nucleon effective mass comes from the momentum-dependent potential
in non-relativistic models (see, e.g., Chapter 3 of
Ref.~\cite{Li08}). The isospin splitting of the in-medium nucleon
effective mass is thus related to the momentum dependence of the
symmetry potential in non-relativistic models~\cite{Li04}. It has
been found that the isospin splitting of the nucleon effective mass
is as important as the nuclear symmetry energy in understanding the
isospin dynamics in nuclear
reactions~\cite{Riz05,Gio10,Fen11,Fen12,Zha14,Xie14,kong15}, and has
ramifications in the thermodynamic properties of isospin asymmetric
nuclear matter as well~\cite{OLi11,xu15}. Moreover, the
neutron-proton effective mass splitting is actually inter-related to
the nuclear symmetry energy through the Hugenholtz-Van Hove
theorem~\cite{XuC10,BAL13}. For a recent review on the isospin
splitting of the nucleon effective mass, we refer the reader to
Ref.~\cite{BAL15}.

Giant resonances of nuclei serve as a useful probe of nuclear
interactions and the EoS of nuclear matter at subsaturation
densities. The studies on giant resonances mainly follow two
methods, i.e., the random-phase approximation and the transport
model calculations. As a breathing oscillation mode in the radial
direction of a nucleus, the isoscalar giant monopole resonance (GMR)
is a good probe of the incompressibility of nuclear
matter~\cite{Bla95,You99,Agr03,Col04,Tod05}, while the isoscalar
giant quadruple resonance (ISGQR), an oscillation mode with
quadruple deformation of a nucleus, has been found to be much
affected by the isoscalar nucleon effective mass
$m_s^*$~\cite{Boh75,Boh79,Bla80,Klu09,Roc13a,zhangzhen16}. On the
other hand, the isovector giant dipole resonance (IVGDR) and the
pygmy dipole resonance (PDR), with the former an oscillation mode
between the centers of mass of neutrons and protons and the latter
that between the neutron skin and the nucleus core, are valuable
probes of the nuclear symmetry energy at subsaturation
densities~\cite{Kli07,Tri08,Car10,Rei10,exp2,Pie12,Bar12,Roc12,Vre12,Roc13b,tao13,Col14,Roc15,zhangzhen15,zhenghua16}.
Since the nuclear symmetry energy acts as a restoring force for the
IVGDR, the main frequency of the IVGDR oscillation, i.e., the centroid
energy $E_{-1}$, is related to $E_{sym}$ at subsaturation densities
or its slope parameter $L$ at the saturation density~\cite{Tri08,tao13,zhenghua16}. The
electric dipole polarizability $\alpha_D$ has a strong correlation
with the neutron skin thickness $\Delta
r_{np}$~\cite{Rei10,zhenghua16}, and $\alpha_D$ times $E_{sym}$ at
the saturation density shows a good linear dependence on
$L$~\cite{Roc13b,Roc15,zhangzhen15}. It was argued that the accurate
knowledge of $\alpha_D$ and $\Delta r_{np}$ can help constrain
significantly the nuclear symmetry energy at subsaturation
densities~\cite{Pie12}. The effect of the neutron-proton effective
mass splitting on the IVGDR properties was realized only
recently~\cite{zhangzhen16}. It is also worth mentioning that the
dynamical isovector dipole collective motion in fusion reactions
could be a probe of the nuclear symmetry energy as well from
transport model
studies~\cite{Bar01a,Bar01b,Pap05,Bar09,wuhongli10,yeshaoqiang13}.

Recently we have improved our isospin- and momentum-dependent
interaction (MDI)~\cite{Das03,Chen05}, which was previously
extensively used in the studies of thermodynamic properties of
nuclear matter, dynamics of nuclear reactions, and properties of
compact stars (see Ref.~\cite{Che14} for a review). In the improved
isospin- and momentum-dependent interaction (ImMDI)~\cite{xu15}, the
momentum dependence of the mean-field potential has been fitted to
that extracted from proton-nucleus scatterings up to the nucleon
kinetic energy of about 1 GeV, and more isovector parameters are
further introduced so that the density dependence of the symmetry
energy and the momentum dependence of the symmetry potential, or
equivalently, the isospin splitting of the nucleon effective mass,
can be mimicked separately. In the present study, we are going to
investigate the effect of the nuclear symmetry energy and the
neutron-proton effective mass splitting on the centroid energy
$E_{-1}$ of IVGDR as well as the electric dipole polarizability
$\alpha_D$, by employing the ImMDI interaction together with the
isospin-dependent Boltzmann-Uehling-Uhlenbeck (IBUU) transport
model. With the experimental data of $E_{-1}$ and $\alpha_D$ from
the IVGDR in $^{208}$Pb available~\cite{exp1,exp2,Roc15}, we are
able to constrain both the slope parameter of the symmetry energy
and the isospin splitting of the nucleon effective mass, once the
isoscalar nucleon effective mass is well determined.
Section~\ref{ImMDI and giant resonance} gives a brief introduction
to the ImMDI interaction and the necessary formalisms for IVGDR.
Detailed studies on the extraction of $m_s^*$ and the constraint on
$L$ and $m_{n-p}^{\ast}$ from IVGDR are presented in
Sec.~\ref{results}. We conclude in Sec.~\ref{summary}.

\section{Theory}
\label{ImMDI and giant resonance}

\subsection{An improved isospin- and momentum-dependent interaction}

The potential energy density functional in the asymmetric nuclear
matter with isospin asymmetry $\delta$ and nucleon number density
$\rho$ from the ImMDI interaction can be expressed
as~\cite{Das03,xu15}
\begin{eqnarray}
V(\rho ,\delta ) &=&\frac{A_{u}\rho _{n}\rho _{p}}{\rho _{0}}+\frac{A_{l}}{%
2\rho _{0}}(\rho _{n}^{2}+\rho _{p}^{2})+\frac{B}{\sigma
+1}\frac{\rho
^{\sigma +1}}{\rho _{0}^{\sigma }}  \notag \\
&\times &(1-x\delta ^{2})+\frac{1}{\rho _{0}}\sum_{\tau ,\tau
^{\prime
}}C_{\tau ,\tau ^{\prime }}  \notag \\
&\times &\int \int d^{3}pd^{3}p^{\prime }\frac{f_{\tau }(\vec{r}, \vec{p}%
)f_{\tau ^{\prime }}(\vec{r}, \vec{p}^{\prime })}{1+(\vec{p}-\vec{p}^{\prime
})^{2}/\Lambda ^{2}}. \label{MDIV}
\end{eqnarray}%
The single-particle potential of a nucleon with isospin $\tau$ and
momentum $\vec{p}$ in the asymmetric nuclear matter with the isospin
asymmetry $\delta$ and nucleon number density $\rho$ can be
expressed as
\begin{eqnarray}
U_\tau(\rho ,\delta ,\vec{p}) &=&A_{u}\frac{\rho _{-\tau }}{\rho _{0}}%
+A_{l}\frac{\rho _{\tau }}{\rho _{0}}  \notag \\
&+&B\left(\frac{\rho }{\rho _{0}}\right)^{\sigma }(1-x\delta ^{2})-4\tau x\frac{B}{%
\sigma +1}\frac{\rho ^{\sigma -1}}{\rho _{0}^{\sigma }}\delta \rho
_{-\tau }
\notag \\
&+&\frac{2C_l}{\rho _{0}}\int d^{3}p^{\prime }\frac{f_{\tau }(%
\vec{r}, \vec{p}^{\prime })}{1+(\vec{p}-\vec{p}^{\prime })^{2}/\Lambda ^{2}}
\notag \\
&+&\frac{2C_u}{\rho _{0}}\int d^{3}p^{\prime }\frac{f_{-\tau }(%
\vec{r}, \vec{p}^{\prime })}{1+(\vec{p}-\vec{p}^{\prime })^{2}/\Lambda ^{2}}.
\label{MDIU}
\end{eqnarray}%
In the above, $\rho _{n}$ and $\rho _{p}$ are the number density of
neutrons and protons, $\rho _{0}$ is the saturation density, and
$\delta =(\rho _{n}-\rho _{p})/\rho$ is the isospin asymmetry.
$f_{\tau }(\vec{r}, \vec{p})$ is the phase-space distribution
function with $\tau=1(-1)$ being the isospin label of neutrons
(protons). The potential energy density of Eq.~(\ref{MDIV}) can be
derived based on the Hartree-Fock calculation from an effective
interaction with a zero-range density-dependent term and a
finite-range Yukawa-type term~\cite{Xu10}.

In the ImMDI interaction~\cite{xu15}, the isovector parameters $x$,
$y$, and $z$ are introduced to vary respectively the slope parameter
of the symmetry energy, the momentum dependence of the symmetry
potential or the neutron-proton effective mass splitting, and the
value of the symmetry energy at the saturation density, via the
following relations
\begin{eqnarray}
A_{l}(x,y)&=&A_{0} + y + x\frac{2B}{\sigma +1},   \label{AlImMDI} \nonumber \\
A_{u}(x,y)&=&A_{0} - y - x\frac{2B}{\sigma +1},   \label{AuImMDI}  \nonumber \\
C_l(y,z)&=&C_{l0} - 2(y-2z)\frac{p^2_{f}}{\Lambda^2\ln [(4 p^2_{f} + \Lambda^2)/\Lambda^2]}, \nonumber \\
C_u(y,z)&=&C_{u0} + 2(y-2z)\frac{p^2_{f}}{\Lambda^2\ln[(4 p^2_{f} + \Lambda^2)/\Lambda^2]}, \nonumber \\
\end{eqnarray}
where $p_{f}=\hbar(3\pi^{2}\rho_0/2)^{1/3}$ is the nucleon Fermi
momentum in symmetric nuclear matter at the saturation density. We
set $z=0$ all through the manuscript. The values of the parameters
$A_0$, $B$, $C_{u0}$, $C_{l0}$, $\sigma$, $\Lambda$, $x$, and $y$
can be fitted or solved from the saturation density $\rho_0$, the
binding energy $E_0$ at $\rho_0$, the incompressibility $K_0$, the
mean-field potential $U_0^\infty$ at infinitely large nucleon
momentum at $\rho_0$, the isoscalar nucleon effective mass $m_s^*$
at $\rho_0$, the symmetry energy $E_{sym}$ and its slope parameter
$L$ at $\rho_0$, and the isovector nucleon effective mass $m_v^*$ at
$\rho_0$, as detailed in \ref{app}.

\subsection{Isovector giant dipole resonance}

The isovector giant dipole resonance (IVGDR) is a collective
vibration of protons against neutrons, with the isovector dipole
operator defined as
\begin{eqnarray}
\hat{D}=\frac{N Z}{A} \hat{X},
\end{eqnarray}
where $N$, $Z$, and $A=N+Z$ are the number of neutrons, protons, and
the mass number of the nucleus, respectively, and $\hat{X}$ is the
distance between the centers of mass of protons and neutrons in the
nucleus. The strength function of IVGDR can be calculated from the
isovector dipole moment via
\begin{eqnarray}\label{se}
S\left ( E \right )=\frac{-Im\left [\tilde{D}(\omega)\right ]}{\pi
\eta },
\end{eqnarray}
where $\tilde{D}(\omega)=\int_{t_{0}}^{t_{max}}D\left ( t \right
)e^{i\omega t}dt$ is the fourier transformation of the isovector
dipole moment with $E=\hbar\omega$. In order to initialize the
isovector dipole oscillation, the initial momenta of protons and
neutrons are given a perturbation in the opposite direction
according to~\cite{urban12}
\begin{eqnarray}
p_{i}\rightarrow\left\{\begin{matrix}
p_{i}-\eta \frac{N}{A}   ~~\rm (protons)\\
p_{i}+\eta \frac{N}{A}   ~~\rm (neutrons)
\end{matrix}\right.,
\end{eqnarray}
with $p_{i}$ being the momentum of the $i$th nucleon along $\hat{X}$, and the perturbation parameter $\eta  =25$ MeV/c used in the
present study. One sees that a larger $\eta$ leads to a larger amplitude of the oscillation $D(t)$, while the strength function $S(E)$ is independent of $\eta$ as the amplitude is cancelled from the denominator. Once the strength function of IVGDR is obtained, the
moments of the strength function can be calculated from
\begin{eqnarray}
m_{k}=\int_0^\infty dE E^{k}S(E).
\end{eqnarray}

Both the centroid energy and the electric dipole polarizability can
be measured from photoabsorption experiments. The centroid energy,
corresponding to the peak energy in the photoabsorption energy
spectrum, is the main frequency of IVGDR. The electric dipole
polarizability, characterizing the response of the nucleus to the
external electric field, is related to the photoabsorption cross
section $\sigma_{f}$ via~\cite{exp2}
\begin{eqnarray}
\alpha _{D}=\frac{\hbar c}{2\pi ^{2}}\int \frac{\sigma _{f}}{\omega ^{2}}d\omega.
\end{eqnarray}
Since both the strength function $S(E)$ and the photoabsorption cross section $\sigma_{f}$ are related to the energy spectrum of the excited states in the nucleus, the centroid energy and the electric dipole polarizability can be expressed in terms of the
moments $m_{k}$ of the strength function as
\begin{eqnarray}
E_{-1}=\sqrt{m_{1}/m_{-1}}
\end{eqnarray}
and
\begin{eqnarray}
\alpha_{D}=2e^{2}m_{-1}.
\end{eqnarray}
We note that the relation between $\alpha_D$ and $m_{-1}$ is
different from that in Ref.~\cite{zhangzhen16}, because we only
consider one-dimensional oscillation, and the definition of the
isovector giant dipole operator is different from that used in
Ref.~\cite{zhangzhen16} based on a random-phase approximation
approach.

In the following calculations, we fit the time evolution of the isovector
giant dipole moment with the function
\begin{eqnarray}\label{fit}
D(t)=a\sin (bt) e^{-ct},
\end{eqnarray}
where $a$, $b$, and $c$ are fitting constants characterizing the
amplitude, the frequency, and the damping time of IVGDR,
respectively. With the help of Eq.~(\ref{fit}), the strength
function $S(E)$, the moments of the strength function $m_{-1}$ and
$m_{1}$, the centroid energy $E_{-1}$ of IVGDR, and the electric
dipole polarizability $\alpha_{D}$ can be expressed analytically in
terms of the fitting constants respectively as
\begin{eqnarray}
S(E)&=&\frac{-ac}{2\pi \eta }\left [ \frac{1}{c^{2}+\left ( b+\frac{E}{\hbar} \right )^{2}} - \frac{1}{c^{2}+\left ( b-\frac{E}{\hbar} \right )^{2}} \right], \notag \\
m_{-1}&=&\frac{-ab}{2\eta\left ( b^{2}+c^{2} \right )}, \notag \\
m_{1}&=&\frac{-ab}{2\eta}, \notag \\
E_{-1}&=&\sqrt{m_{1}/m_{-1}}= \sqrt{b^{2}+c^{2}}, \notag \\
\alpha_{D}&=&2e^{2}m_{-1}=\frac{-2e^{2}ab}{2\eta\left ( b^{2}+c^{2} \right )}.
\label{abc}
\end{eqnarray}

\section{Results and discussions}
\label{results}

In the following study, we employ the IBUU transport model together
with the ImMDI interaction to investigate the giant resonances of
nuclei. The positions of the projectile and target in the IBUU
transport model are fixed, i.e., with zero beam energy. The initial
density distribution is sampled according to that generated from
Skyrme-Hartree-Fock calculations with the same physics quantities used in the ImMDI interaction,
such as $L$, $m_v^*$, etc., listed in Table~\ref{para}. The initial
nucleon momentum is sampled accordingly to the local density from
the Thomas-Fermi approximation. We generate events from 40 runs with
each run 200 test particles. Since the oscillation generally lasts
for hunderds of fm/c, in order to improve the stability in the
calculation with the momentum-dependent mean-field potential, we use
the nuclear matter approximation in the real calculation by taking
the phase-space distribution function as $f_{\tau}(\vec{r},
\vec{p})=\frac{2}{h^{3}}\Theta\left(p_{f\tau}-p\right)$ and using
the analytical expression (Eqs.~(\ref{MDIU}) and (\ref{int1})) for
the momentum-dependent mean-field potential. This is similar to the
spirit of the Thomas-Fermi approximation in the case that the
vibration compared to the stable distribution is small. With this
treatment, we found that up to $t=500$ fm/c only about $17\%$ of the
total nucleons become free particles.

\subsection{Extract the isoscalar nucleon effective mass}

\begin{figure}[h]
\centerline{\includegraphics[scale=0.8]{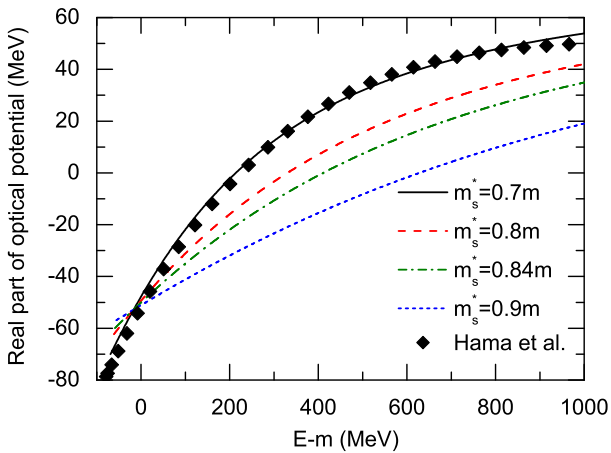}} \caption{(Color
online) Single-particle potentials in symmetric nuclear nuclear
matter at the saturation density as a function of the nucleon energy
subtracted by the nucleon rest mass from the ImMDI interaction with
different isoscalar nucleon effective mass $m_{s}^{*}$. The real
part of the optical potential extracted from proton-nucleus
scatterings by Hama et al.~\cite{hama1990,cooper1993} is shown by
scatters for comparison. } \label{F1}
\end{figure}

We first extract the isoscalar nucleon effective mass from the
optical potential. The single-particle potentials in symmetric
nuclear nuclear matter at the saturation density, with different
values of the isoscalar nucleon effective mass $m_{s}^{*}$ but same
other isoscalar properties of the nuclear interaction, are displayed
in Fig.~\ref{F1}. We can see that only the parameter set that leads
to $m_{s}^{*}=0.7 m$, with $m$ being the nucleon mass in free space,
can describe reasonably well the real part of the optical potential
extracted from proton-nucleus scatterings by Hama et
al.~\cite{hama1990,cooper1993}.

\begin{table*}\small
  \caption{The physics quantities and the corresponding parameters for the ImMDI interaction as well as the results from the IVGDR in $^{208}$Pb.}
    \begin{tabular}{|c || c|c|c|| c|c|c|}
    \hline
      & Set I(a) & Set I(b) & Set I(c) & Set II(a) & Set II(b) & Set II(c) \\
   \hline
    $A_0$ (MeV)     & -66.963 & -66.963 & -66.963 & -96.799 & -96.799 & -96.799 \\
   \hline
    $B$ (MeV)     & 141.963 & 141.963 & 141.963 & 171.799 & 171.799 & 171.799  \\
   \hline
    $C_{u0}$ (MeV)    & -99.70 & -99.70 & -99.70 & -90.19 & -90.19 & -90.19 \\
   \hline
    $C_{l0}$ (MeV) & -60.49 & -60.49 & -60.49 & -50.03 & -50.03 & -50.03 \\
   \hline
    $\sigma$         & 1.2652 & 1.2652 & 1.2652 & 1.2704 & 1.2704 & 1.2704 \\
   \hline
    $\Lambda$ ($p_{f}$)    & 2.424 & 2.424 & 2.424 & 3.984 & 3.984 & 3.984 \\
   \hline
    $x$           & 0 & 1 & 1 & 0 & 1 & 1 \\
   \hline
    $y$ (MeV)     & -115 & -115 & 115 & -115 & -115 & 115  \\
   \hline
   \hline
     $\rho_0$ (fm$^{-3}$) & 0.16 & 0.16 & 0.16 & 0.16 & 0.16 & 0.16  \\
   \hline
     $E_0(\rho_0)$ (MeV) & -16 & -16 & -16 & -16 & -16 & -16  \\
   \hline
     $K_0$ (MeV)  & 230 & 230 & 230 & 230 & 230 & 230  \\
   \hline
     $U_0^\infty$ (MeV)   & 75 & 75 & 75 & 75 & 75 & 75  \\
   \hline
     $m_s^*$ ($m$)  & 0.7 & 0.7 & 0.7 & 0.84 & 0.84 & 0.84  \\
   \hline
     $E_{sym}(\rho_0)$ (MeV)   & 32.5 & 32.5 & 32.5 & 32.5 & 32.5 & 32.5  \\
   \hline
     $L$ (MeV)        & 58.57 & 8.70 & 60.00 & 72.63 & 11.24 & 36.03  \\
   \hline
     $m_v^*$ ($m$)   & 0.537 & 0.537 & 0.853 & 0.712 & 0.712 & 0.928   \\
   \hline
   \hline
    $a$ (fm) & -28.13$\pm$0.33 & -27.16$\pm$0.25 & -19.55$\pm$0.14 & -24.30$\pm$0.21 & -22.69$\pm$0.21 & -19.34$\pm$0.16 \\
   \hline
    $b$ (fm$^{-1}$)  & 0.0782$\pm$0.0001 & 0.0883$\pm$0.0001 & 0.0614$\pm$0.0001 & 0.0662$\pm$0.0001 & 0.0760$\pm$0.0001 & 0.0640$\pm$0.0001\\
   \hline
    $c$ (fm$^{-1}$)  & 0.0075$\pm$0.0001 & 0.0099$\pm$0.0001 & 0.0040$\pm$0.0001 & 0.0052$\pm$0.0001 & 0.0075$\pm$0.0001 & 0.0054$\pm$0.0001 \\
   \hline
   $E_{-1}$ (MeV)  & 15.50$\pm$0.0001 & 17.53$\pm$0.0001 & 12.13$\pm$0.0001 & 13.10$\pm$0.0001 & 15.06$\pm$0.0001 & 12.68$\pm$0.0001 \\
   \hline
   $\alpha_D$ (fm$^3$) & 20.5$\pm$0.2 & 17.5$\pm$0.2 & 18.3$\pm$0.1 & 21.0$\pm$0.2 & 17.0$\pm$0.2 & 17.3$\pm$0.1\\
   \hline
    \end{tabular}
  \label{para}
\end{table*}

\begin{figure}[h]
\centerline{\includegraphics[scale=0.3]{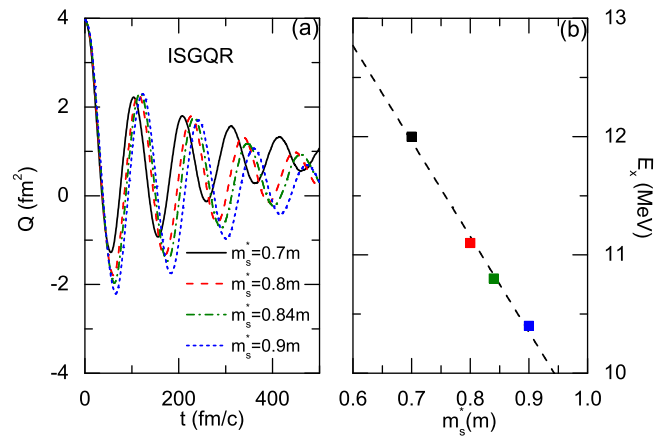}} \caption{(Color
online) The time evolution of the isoscalar giant quadruple moment
in $^{208}$Pb (a) and the correlation between the corresponding
excitation energy $E_{x}$ and the isoscalar nucleon effective mass
$m_{s}^{\ast}$ (b).} \label{F4}
\end{figure}

Next, we extract the value of $m_{s}^{\ast }$ from the isoscalar
giant quadruple resonance (ISGQR) in $^{208}$Pb, with the operator
written as
\begin{eqnarray}
\hat{Q}=\sum_{i=1}^{A} r_{i}^{2} Y_{20} \left ( \hat{r_{i}} \right
)= \sum_{i=1}^{A} \sqrt{\frac{5}{16\pi }}\left (
3z_{i}^{2}-r_{i}^{2} \right ),
\end{eqnarray}
where $r_i$ and $z_i$ are respectively the radial and $z$-direction
coordinate of the $i$th nucleon, and $Y_{20}$ is the spherical
harmonic function. Noticing that the following scaling relation in
the ISGQR is observed (see, e.g., Ref.~\cite{GQRla02})
\begin{eqnarray}
\left\{\begin{matrix}x\rightarrow x / \lambda  \\
y\rightarrow y / \lambda
\\ z\rightarrow \lambda ^{2} z \end{matrix}\right.
\left\{\begin{matrix}p_{x}\rightarrow  \lambda p_{x} \\
p_{y}\rightarrow \lambda p_{y}
\\ p_{z}\rightarrow  p_{z} /\lambda ^{2} \end{matrix}\right.,
\end{eqnarray}
we choose $\lambda=1.1$ in our simulation to initialize the
oscillation. The value of $\lambda$ is close to $1$ corresponding to
a small vibration with respect to the equilibrium distribution.
Again, we found that the value of $\lambda$ only affects the
amplitude but has almost no effect on the frequency of ISGQR. The
time evolution of the isoscalar giant quadruple moment, with
different values of the isoscalar nucleon effective mass but same
other isoscalar properties of nuclear interaction, is displayed in
the left panel of Fig.~\ref{F4}. It is seen that a smaller isoscalar
nucleon effective mass $m_{s}^{\ast}$ leads to a larger frequency of
the oscillation. From the Fourier transformation of $Q(t)$, the
linear correlation between the isoscalar nucleon effective mass
$m_{s}^{\ast}$ and the excitation energy of ISGQR is observed in the
right panel of Fig.~\ref{F4}. It is found that the result from the
isoscalar nucleon effective mass $m_{s}^{\ast}=0.84m$ reproduces
best the excitation energy $E_{x}=10.9\pm 0.1$ MeV of ISGQR in
$^{208}$Pb extracted from $\alpha$-nucleus scattering
experiments~\cite{exp3}.

The different values of the isoscalar effective mass $m_{s}^{*}$
extracted from the optical potential and from the excitation energy
of ISGQR in $^{208}$Pb represent the theoretical uncertainties in
the present study. In the following study, we will thus employ the
parameter Set I and Set II that lead respectively to $m_{s}^{*}=0.7
m$ and $0.84 m$ together with different combinations of the
isovector parameters $x$ and $y$ to study the isovector giant dipole
resonances of nuclei. The values of the parameters and the
corresponding physics quantities are detailed in Table~\ref{para}.

\subsection{Constrain the symmetry energy and the neutron-proton effective mass splitting}

\begin{figure}[h]
\centerline{\includegraphics[scale=0.8]{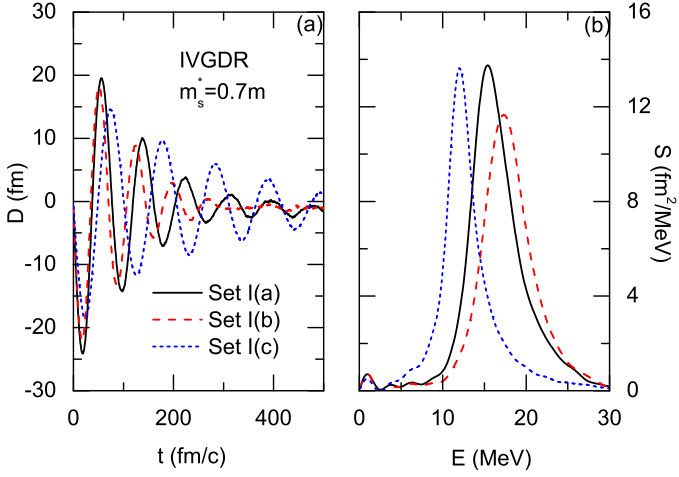}} \caption{(Color
online) The time evolution of the isovector giant dipole moment (a)
and the corresponding strength function (b) in $^{208}$Pb with
$m_{s}^{\ast }=0.7m$.} \label{F2}
\end{figure}

We first employ the parameter Set I with $m_{s}^{\ast }=0.7m$, which
is able to reproduce the empirical optical potential as shown in
Fig.~\ref{F1}, to study the properties of the IVGDR in $^{208}$Pb.
Three different parameter sets, with different combinations of the
slope parameter $L$ of the symmetry energy and the isovector
effective mass $m_v^*$ detailed as Set I(a), Set I (b), and Set I(c)
in Table~\ref{para}, are employed in the study. The initial momenta
of neutrons and protons are modified with the perturbation parameter
$\eta$ as detailed in Sec.~\ref{ImMDI and giant resonance}B. The
oscillation amplitude is proportional to $\eta$ while its frequency
is found to be insensitive to the choice of $\eta$. The resulting
time evolution of the isovector dipole moment is displayed in the
left panel of Fig.~\ref{F2}. One sees that the time evolution of
$D(t)$ follows a good damping oscillation mode as in
Eq.~(\ref{fit}). In order to avoid oscillation in the Fourier
transformation due to the finite $t_{max}$ in the transport model
calculation, a damping factor of $\exp(-\frac{\gamma t}{2 \hbar})$
with the width $\gamma =2$ MeV is multiplied to the isovector dipole
moment $D(t)$ in calculating the strength function $S(E)$ from
Eq.~(\ref{se}) as in Ref.~\cite{urban12}. This slightly affects the
damping coefficient $c$ in Eq.~(\ref{fit}) but has very small
effects on the final results as can be seen from the analytical
formulaes of Eq.~(\ref{abc}). The resulting strength functions from
the numerical Fourier transformation is displayed in the right panel
of Fig.~\ref{F2}. It is seen that with a softer symmetry energy (Set
I(b)), the main frequency, i.e., the centroid energy in the IVGDR,
is larger, due to the larger symmetry energy at subsaturation
densities acting as a stronger restoring force. On the other hand,
the centroid energy is sensitive to the isovector nucleon effective
mass as well, with a larger isovector effective mass (Set I(c))
leading to a smaller centroid energy of IVGDR. This could be
understood since the oscillation frequency is smaller with a larger
reduced mass of the system, with the latter attributed to the larger
isovector effective mass once the isoscalar effective mass is fixed.

\begin{figure}[h]
\centerline{\includegraphics[scale=0.35]{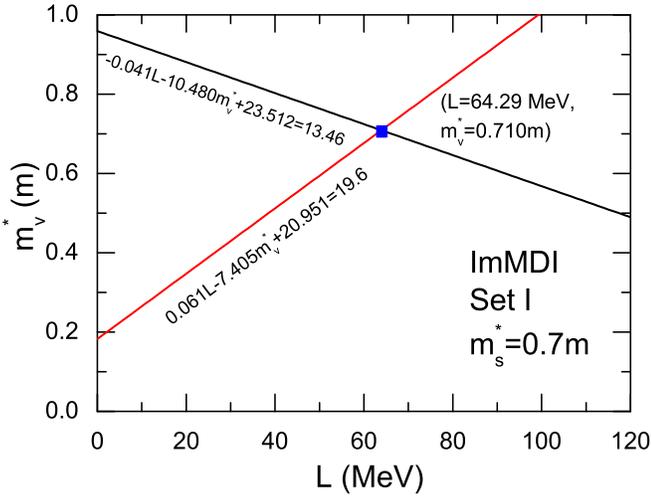}} \caption{(Color
online) The linear relations between the symmetry energy slope
parameter $L$ as well as the isovector nucleon effective mass
$m_{v}^{\ast}$ and the centroid energy $E_{-1}$ as well as the
electric dipole polarizability $\alpha_D$ from the IVGDR in
$^{208}$Pb, respectively, with the isoscalar nucleon effective mass
$m_{s}^{\ast }=0.7m$.} \label{F3}
\end{figure}

The constants $a$, $b$, and $c$ can be obtained from fitting the
isovector dipole moment $D(t)$ according to Eq.~(\ref{fit}), and
their detailed values for the three parameter sets are listed in
Table~\ref{para}. The resulting centroid energy $E_{-1}$ and the
electric dipole polarizability $\alpha_{D}$ calculated analytical
according to Eq.~(\ref{abc}) are also listed in Table~\ref{para}. It
is seen that for a given isovector effective mass $m_{v}^{\ast}$, a
larger slope parameter $L$ leads to a smaller centroid energy
$E_{-1}$ and a larger electric dipole polarizability $\alpha _{D}$.
Analogously, for a given slope parameter $L$, a larger isovector
effective mass $m_{v}^{\ast}$ leads to a smaller centroid energy
$E_{-1}$ and a smaller electric dipole polarizability $\alpha _{D}$.
The rigorous tool to treat such a problem of multiple experimental
results with multiple parameters is the Bayesian
framework~\cite{Hab07,Goz12,Pra15,San16}, which is beyond the scope
of the present study. On the other hand, we found both $E_{-1}$ and
$\alpha _{D}$ are linearly correlated with $L$ and $m_{v}^{\ast}$,
and the corresponding relations based on our transport calculations
turns out to be
\begin{eqnarray}\label{eq1}
\left\{\begin{matrix} -0.041L-10.480m_{v}^{*}+23.512=E_{-1},\\
0.061L-7.405m_{v}^{*}+20.951=\alpha_D,
\end{matrix}\right.
\end{eqnarray}
with $L$ in MeV, $m_v^*$ in $m$, $E_{-1}$ in MeV, and $\alpha_D$ in
fm$^3$. The centroid energy of the IVGDR in $^{208}$Pb obtained
experimentally from the photoabsorption measurement is
$E_{-1}=13.46$ MeV~\cite{exp1}, while the electric dipole
polarizability is $\alpha _{D}=19.6\pm 0.6$ fm$^{3}$ measured from
photoabsorption cross section by Tamii et al.~\cite{exp2} and with
further correction by subtracting quasideuteron
excitations~\cite{Roc15}. With the above experimental data
available, the constraints on the slope parameter $L$ and the
isovector effective mass can be solved from Eq.~(\ref{eq1}) as
\begin{eqnarray}\label{lmv1}
L&=&64.29\pm11.84 ~(\rm MeV), \\ m_{v}^{\ast}/m&=&0.710\pm0.046 ,
\end{eqnarray}
where the error bars, which originate from the fitting and the
statistical error of $a$, $b$, and $c$ listed in Table~\ref{para},
are calculated from the error transfer. We found that the ImMDI parameterization
with the mean values of $L$ and $m_v^*$ in Eq.~(\ref{lmv1}) gives
very close results of $E_{-1}$ and $\alpha_D$ compared with the
experimental data, justifying the linear relation of Eq.~(\ref{eq1}).
The corresponding isospin splitting of the nucleon effective mass
deduced from Eq.~(\ref{mnp}) is
\begin{eqnarray}
 (m_{n}^{\ast}-m_{p}^{\ast})/m=  (-0.019 \pm 0.090)\delta.
\end{eqnarray}
The above constraint is, however, different from that obtained by
analyzing nucleon-nucleus scattering data within an
isospin-dependent optical model~\cite{Li15}.

\begin{figure}[h]
\centerline{\includegraphics[scale=0.8]{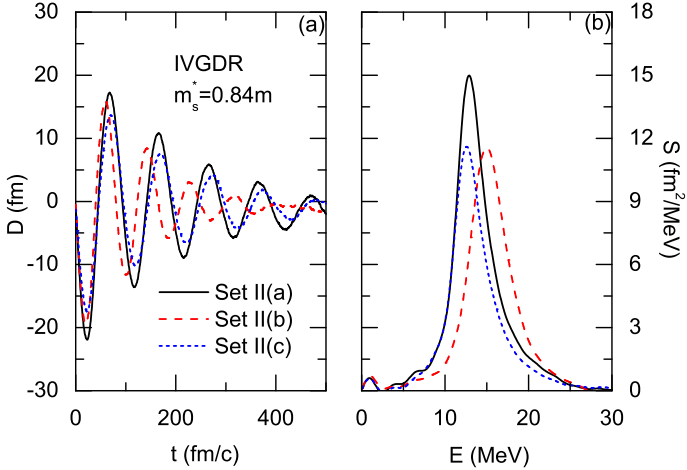}} \caption{(Color
online) Same as Fig.~\ref{F2} but with the isoscalar nucleon
effective mass $m_{s}^{\ast }=0.84m$.} \label{F5}
\end{figure}

Next, we choose the isoscalar nucleon effective mass to be
$m_{s}^{\ast}=0.84m$ while keeping other physics quantities the same
in the ImMDI parameterization, and the resulting parameter sets are
listed as Set II(a), Set II(b), and Set II(c) in Table~\ref{para}
with different combinations of the slope parameter $L$ of the
symmetry energy and the isovector nucleon effective mass
$m_{v}^{\ast}$. With the same calculation method, the time evolution
of the isovector dipole moment in $^{208}$Pb and the corresponding
strength function are displayed respectively in the left and right
panel of Fig.~\ref{F5}. With $a$, $b$, and $c$ fitted by
Eq.~(\ref{fit}), and the analytical results of the centroid energy
$E_{-1}$ and the electric dipole polarizability $\alpha_D$ obtained
according to Eq.~(\ref{abc}), we can get the similar linear
relation from transport calculations as
\begin{eqnarray}\label{eq2} \left\{\begin{matrix}
-0.032L-7.346m_{v}^{*}+20.651=E_{-1},\\
0.065L-6.368m_{v}^{*}+20.845=\alpha_D.
\end{matrix}\right.
\end{eqnarray}
With the available experimental data of $E_{-1}$ and $\alpha_D$, the
slope parameter $L$ of the symmetry energy and the isovector nucleon
effective mass from the constraint of the ISQGR and the IVGDR in
$^{208}$Pb are
\begin{eqnarray}\label{lmv2}
L&=&53.85\pm10.29 ~(\rm MeV), \\ m_{v}^{\ast}/m&=&0.744\pm0.045.
\end{eqnarray}
Again, we found that the mean values of $L$ and $m_v^*$ reproduce
very well the experimental results of $E_{-1}$ and $\alpha_D$ within
the statistical error based on our transport model calculations,
justifying the linear relation of Eq.~(\ref{eq2}). The corresponding
neutron-proton effective mass splitting deduced from Eq.~(\ref{mnp})
is thus
\begin{eqnarray}
 (m_{n}^{\ast}-m_{p}^{\ast})/m= (0.216 \pm 0.114) \delta.
\end{eqnarray}
The constraint from both ISGQR and IVGDR on the neutron-proton
effective mass splitting is consistent with that obtained in
Ref.~\cite{Li15}.

\begin{figure}[h]
\centerline{\includegraphics[scale=0.35]{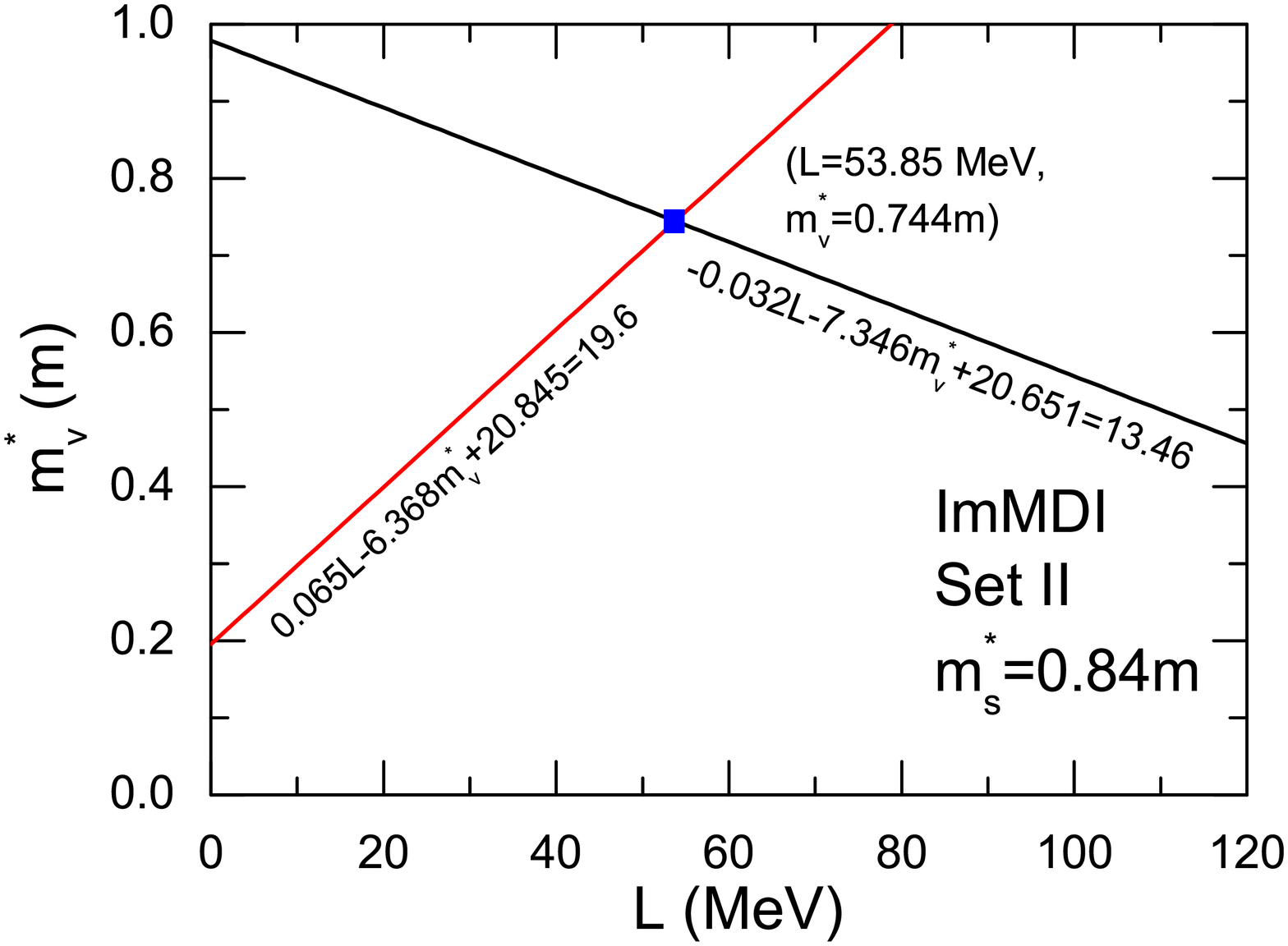}} \caption{(Color
online) Same as Fig.~\ref{F3} but with the isoscalar nucleon
effective mass $m_{s}^{\ast }=0.84m$.} \label{F6}
\end{figure}

\section{Conclusions}
\label{summary}

Based on an improved isospin- and momentum-dependent interaction
(ImMDI) and an isospin-dependent Boltzmann-Uehling-Uhlenbeck (IBUU)
transport model, we have studied the effect of the slope parameter
$L$ of the nuclear symmetry energy and the isovector nucleon
effective mass $m_{v}^{*}$ on the centroid energy $E_{-1}$ of the
isovector giant dipole resonance (IVGDR) and the electric dipole
polarizability $\alpha_{D}$ in $^{208}$Pb. We found that both
$E_{-1}$ and $\alpha_D$ are almost linearly correlated with $L$ and
$m_{v}^{*}$. With a given isoscalar nucleon effective mass, we are
able to constrain the values of $L$ and $m_v^*$ with the available
experimental data of $E_{-1}$ and $\alpha_D$. From the isoscalar
nucleon effective mass $m_s^*=0.7m$ constrained by the empirical
optical potential, we obtain a constraint of $L=64.29\pm11.84 (\rm
MeV)$ and $m_{v}^{\ast}/m=0.710\pm0.046$, resulting in the isospin
splitting of the nucleon effective mass within
$(m_{n}^{\ast}-m_{p}^{\ast})/m= (-0.019 \pm 0.090)\delta$, with
$\delta$ being the isospin asymmetry of nuclear medium. From the
isoscalar nucleon effective mass $m_s^*=0.84m$ extracted from the
excitation energy of the isoscalar giant quadruple resonance (ISGQR)
in $^{208}$Pb, we obtain a constraint of $L=53.85\pm10.29 (\rm MeV)$
and $m_{v}^{\ast}/m=0.744\pm0.045$, resulting in the isospin
splitting of nucleon effective mass within
$(m_{n}^{\ast}-m_{p}^{\ast})/m= (0.216 \pm 0.114)\delta$. The
constraint on the neutron-proton effective mass splitting from both
ISGQR and IVGDR in $^{208}$Pb is consistent with that from optical
model analyses of nucleon-nucleus elastic scatterings. The
uncertainty of the isoscalar nucleon effective mass has hampered our
accurate constraint on the neutron-proton effective mass splitting.

\begin{acknowledgments}
We thank Chen Zhong for maintaining the high-quality performance of
the computer facility, and acknowledge helpful communications with
Zhen Zhang. JX acknowledges support from the Major State Basic
Research Development Program (973 Program) of China under Contract
No. 2015CB856904 and No. 2014CB845401, the National Natural Science
Foundation of China under Grant No. 11475243 and No. 11421505, the
"100-Talent Plan" of Shanghai Institute of Applied Physics under
Grant No. Y290061011 and No. Y526011011 from the Chinese Academy of
Sciences, the Shanghai Key Laboratory of Particle Physics and
Cosmology under Grant No. 15DZ2272100, and the "Shanghai Pujiang
Program" under Grant No. 13PJ1410600. LWC acknowledges the Major
State Basic Research Development Program (973 Program) in China
under Contract No. 2013CB834405 and No. 2015CB856904, the National
Natural Science Foundation of China under Grant No. 11275125 and No.
11135011, the "Shu Guang" project supported by Shanghai Municipal
Education Commission and Shanghai Education Development Foundation,
the Program for Professor of Special Appointment (Eastern Scholar)
at Shanghai Institutions of Higher Learning, and the Science and
Technology Commission of Shanghai Municipality (11DZ2260700). BAL
acknowledges the National Natural Science Foundation of China under
Grant No. 11320101004, the U.S. Department of Energy, Office of
Science, under Award Number de-sc0013702, and the CUSTIPEN
(China-U.S. Theory Institute for Physics with Exotic Nuclei) under
the US Department of Energy Grant No. DEFG02- 13ER42025. YGM
acknowledges the Major State Basic Research Development Program (973
Program) of China under Contract No. 2014CB845401 and the National
Natural Science Foundation of China under Contract Nos. 11421505 and
11220101005.

\end{acknowledgments}

\appendices

\renewcommand\thesection{APPENDIX~\Alph{section}}
\renewcommand\theequation{\Alph{section}.\arabic{equation}}
\section{Expressions for physics quantities from the ImMDI interaction}
\label{app}

At zero temperature, the phase-space distribution function can be
written as $f_{\tau}(\vec{r},
\vec{p})=\frac{2}{h^{3}}\Theta\left(p_{f\tau}-p\right)$, with
$p_{f\tau}=\hbar(3\pi^{2}\rho_\tau)^{1/3}$ being the Fermi momentum
of nucleons with the isospin label $\tau$, and the
momentum-dependent part of the single-particle potential as well as
that in the potential energy density can be integrated analytically
as~\cite{xu09}
\begin{eqnarray}\label{int1}
&&\int d^{3}p^{\prime }\frac{f_{\tau }(\vec{r}, \vec{p}^{\prime
})}{1+(\vec{p}-\vec{p}^{\prime })^{2}/\Lambda ^{2}} \nonumber \\ =
&&\frac{2}{h^{3}}\pi\Lambda^{3}\left\{\frac{p_{f\tau}^{2}+\Lambda
^{2}-p^{2}}{2p\Lambda}\ln\left[\frac{(p+p_{f\tau})^{2}+\Lambda
^{2}}{(p-p_{f\tau })^{2}+\Lambda ^{2}}\right] \right. \nonumber \\ +
&&\left.\frac{2p_{f\tau} }{\Lambda }-2\left (\arctan\frac{
p+p_{f\tau  }}{\Lambda }-\arctan\frac{ p-p_{f\tau }}{\Lambda }
\right)\right\}\notag\\
\end{eqnarray}
and
\begin{eqnarray}\label{int2}
&&\int \int d^{3}pd^{3}p^{\prime }\frac{f_{\tau }(\vec{r}, \vec{p})f_{\tau ^{\prime }}(\vec{r}, \vec{p}^{\prime })}{1+(\vec{p}-\vec{p}^{\prime})^{2}/\Lambda ^{2}}\nonumber \\ =
&&\frac{1}{6}\left(\frac{4 \pi }{h^{3}}\right)^{2}\Lambda ^{2}\{p_{f}(\tau)p_{f}(\tau ^{\prime})[3(p_{f\tau}^{2}+p_{f\tau ^{\prime}}^{2})-\Lambda ^{2}] \nonumber\\ + &&4\Lambda\left[(p_{f\tau}^{3}-p_{f\tau ^{\prime}}^{3})\arctan\left(\frac{p_{f\tau}-p_{f\tau ^{\prime}}}{\Lambda}\right)\right.\nonumber \\ -
&&\left.(p_{f\tau}^{3}+p_{f\tau ^{\prime}}^{3})\arctan\left(\frac{p_{f\tau}+p_{f\tau ^{\prime}}}{\Lambda}\right)\right] \nonumber\\ +
&&\frac{1}{4}[\Lambda^{4}+6\Lambda^{2}(p_{f\tau}^{2}+p_{f\tau ^{\prime}}^{2})-3(p_{f\tau}^{2}-p_{f\tau ^{\prime}}^{2})^{2}] \nonumber\\ \times
&&\ln\left[\frac{(p_{f\tau }+p_{f\tau ^{\prime}})^{2}+\Lambda ^{2}}{(p_{f\tau }-p_{f\tau ^{\prime}})^{2}+\Lambda ^{2}}\right]\}.
\end{eqnarray}
The binding energy per nucleon for asymmetric nuclear matter can be
expressed as
\begin{eqnarray}
E\left ( \rho ,\delta  \right )=\frac{V\left ( \rho ,T=0,\delta  \right )}{\rho }+E_{k}\left ( \rho ,\delta  \right )
\end{eqnarray}
with the kinetic energy per nucleon calculated from
\begin{eqnarray}
E_{k}\left ( \rho ,\delta  \right )&=&\frac{1}{\rho }\int
d^{3}p\left [ \frac{p^{2}}{2 m} f_{n}\left ( \vec{r}, \vec{p}\right
)+ \frac{p^{2}}{2 m} f_{p}\left ( \vec{r}, \vec{p}\right ) \right ]
\notag\\ &=&\frac{4\pi }{5 m h^{3} \rho }\left (
p_{fn}^{5}+p_{fp}^{5} \right ),
\end{eqnarray}
where $p_{fn\left ( p \right )}=\hbar\left ( 3\pi ^{2}\rho _{n\left
( p \right )} \right )^{\frac{1}{3}}$ is the Fermi momentum of
neutrons (protons), and $m$ is the nucleon mass.

By setting $\rho_{n}=\rho_{p}=\frac{\rho}{2}$ and
$p_{fn}=p_{fp}=p_{f}$, we can express the binding energy per nucleon
for symmetric nuclear matter as~\cite{xu09}
\begin{eqnarray}\label{E0}
&&E_{0}(\rho)\notag\\
&=&\frac{8\pi}{5mh^{3}\rho}p_{f}^{5}+\frac{\rho}{4\rho_{0}}(A_{l}+A_{u})+\frac{B}{\sigma+1}\left(\frac{\rho}{\rho_{0}}\right)^{\sigma}
\notag\\ &+& \frac{1}{3\rho_{0}\rho}(C_{l}+C_{u})\left(\frac{4\pi}{h^{3}}\right)^{2}\Lambda^{2}
\notag\\ &\times& \left[p_{f}^{2}(6p_{f}^{2}-\Lambda^{2})-8 \Lambda p_{f}^{3}\arctan\left(\frac{2p_{f}}{\Lambda}\right)\right.
\notag\\ &+& \left.\frac{1}{4}(\Lambda^{4}+12\Lambda^{2}p_{f}^{2})\ln\left(\frac{4p_{f}^{2}+\Lambda^{2}}{\Lambda^{2}}\right)\right].
\end{eqnarray}
The saturation density is determined by the zero point of the first-order derivative of the binding energy per nucleon, with the latter expressed as~\cite{xu09}
\begin{eqnarray}\label{dE0}
&&\frac{dE_{0}(\rho)}{d\rho}\notag\\
&=&\frac{16\pi}{15mh^{3}\rho^{2}}p_{f}^{5}+\frac{1}{4\rho_{0}}(A_{l}+A_{u})+\frac{B\sigma}{\sigma+1}\frac{\rho^{\sigma-1}}{\rho_{0}^{\sigma}}
\notag \\ &+& \frac{1}{3\rho_{0}\rho^{2}}(C_{l}+C_{u})\left(\frac{4\pi}{h^{3}}\right)^{2}\Lambda^{2}
\\ &\times& \left[2p_{f}^{4}+\Lambda^{2}p_{f}^{2}-\frac{1}{4}(\Lambda^{4}+4\Lambda^{2}p_{f}^{2})\ln\left(\frac{4p_{f}^{2}+\Lambda^{2}}{\Lambda^{2}}\right)\right].\notag
\end{eqnarray}
The incompressibility of symmetric nuclear matter is defined as $K_{0}=9\rho_{0}^{2}(d^{2}E_{0}/d\rho^{2})_{\rho=\rho_{0}}$, with the second-order derivative of the binding energy per nucleon expressed as~\cite{xu09}
\begin{eqnarray}\label{d2E0}
&&\frac{d^{2}E_{0}(\rho)}{d\rho^{2}} \notag\\
&=& -\frac{16\pi}{45 m h^{3} \rho^{3}} p_{f}^{5} + \frac{B \sigma (\sigma-1)}{\sigma+1}\frac{\rho^{\sigma-2}}{\rho_{0}^{\sigma}}
\notag \\ &+& \frac{2}{3\rho_{0}\rho^{3}}(C_{l}+C_{u})\left(\frac{4\pi}{h^{3}}\right)^{2}\Lambda^{2}
 \\ &\times& \left[-\frac{2}{3}p_{f}^{4}-\Lambda^{2}p_{f}^{2}+\Lambda^{2}\left(\frac{\Lambda^{2}}{4}+\frac{2}{3}p_{f}^{2}\right)\ln\left(\frac{4p_{f}^{2}+\Lambda^{2}}{\Lambda^{2}}\right)\right].\notag              \end{eqnarray}
The symmetry energy by definition can be written as~\cite{xu09}
\begin{eqnarray}\label{Esym}
&&E_{sym}(\rho)\notag\\
&=&\frac{1}{2}\left (\frac{ \partial ^{2}E}{\partial \delta ^{2}} \right )_{\delta = 0}
\notag \\ &=&
\frac{8\pi }{9mh^{3}\rho }p_{f}^{5} +\frac{\rho }{4\rho _{0}}\left ( A_{l}-A_{u} \right )-\frac{Bx}{\sigma +1}\left ( \frac{\rho }{\rho _{0}}
\right)^{\sigma } \notag \\ &+&
\frac{C_{l}}{9\rho _{0}\rho }\left ( \frac{4\pi }{h^{3}} \right )^{2}\Lambda ^{2}\left [ 4p_{f}^{4}-\Lambda ^{2}p_{f}^{2}\ln\left(\frac{4p_{f}^{2}+\Lambda ^{2}}{\Lambda ^{2}}\right) \right ]  \\ &+&
\frac{C_{u}}{9\rho _{0}\rho }\left ( \frac{4\pi }{h^{3}} \right )^{2}\Lambda ^{2}\left [ 4p_{f}^{4}-p_{f}^{2}\left(4p_{f}^{2}+\Lambda ^{2}\right )\ln\left(\frac{4p_{f}^{2}+\Lambda ^{2}}{\Lambda ^{2}} \right)\right ]\nonumber.
\end{eqnarray}
The slope parameter of the symmetry energy at the saturation density
is defined as $L=3\rho _{0}[d E_{sym}( \rho)/d \rho]_{\rho =\rho
_{0}}$, with the first-order derivative of the symmetry energy
expressed as~\cite{xu09}
\begin{eqnarray}\label{dEsym}
&&\frac{dE_{sym}(\rho)}{d\rho} \notag\\
&=& \frac{16\pi}{27mh^3\rho^2}p_f^5 + \frac{1}{4\rho_0}(A_l-A_u) \notag\\
&-&\frac{Bx\sigma}{\sigma+1}\frac{\rho^{\sigma-1}}{\rho_0^\sigma} + \frac{C_l+C_u}{27\rho_0\rho^2}\left( \frac{4\pi}{h^3}\right)^2\Lambda^2 \\
&\times&\left[ 4p_f^4 + \Lambda^2p_f^2 \ln \left( \frac{4p_f^2+\Lambda^2}{\Lambda^2}\right)-\frac{8\Lambda^2p_f^4}{4p_f^2+\Lambda^2}\right] \notag\\
&-&\frac{4C_u}{27\rho_0\rho^2}\left( \frac{4\pi}{h^3}\right)^2 \Lambda^2 p_f^4 \left[ \ln \left( \frac{4p_f^2+\Lambda^2}{\Lambda^2}\right)+\frac{8p_f^2}{4p_f^2+\Lambda^2}\right].\notag
\end{eqnarray}

The isoscalar nucleon effective mass $m_{s}^{*}$ is defined as the
nucleon effective mass in symmetric nuclear matter at the saturation
density, and it can be calculated from the mean-field potential
$U_0$ in symmetric nuclear matter via
\begin{eqnarray}\label{ms}
m_{s}^{*}=m\left( 1+\frac{m}{p}\frac{dU_0}{dp}\right) ^{-1}_{p=p_f}.
\end{eqnarray}
In asymmetric nuclear matter, the isovector nucleon effective mass
can be calculated through the following relation
\begin{eqnarray}\label{mv}
\frac{\hbar^{2}}{2 m_{n\left ( p \right )}^{\ast }}=\frac{2\rho
_{n\left ( p \right )}}{\rho _{0}}\frac{\hbar^{2}}{2 {m_{s}^{\ast
}}^{2}}+\left ( 1-\frac{2\rho _{n\left ( p  \right )}}{\rho
_{0}}\right )\frac{\hbar^{2}}{2 m_{v}^{\ast }}, \label{mvmsmnmp}
\end{eqnarray}
with the neutron (proton) effective mass defined as
\begin{eqnarray}\label{ms}
m_{n(p)}^{*}=m\left( 1+\frac{m}{p}\frac{dU_{n(p)}}{dp}\right)
^{-1}_{p=p_f}.
\end{eqnarray}
Keeping the first-order term of the isospin asymmetry $\delta$, the
neutron-proton effective mass splitting is related to the isoscalar
and isovector effective mass through the following relation
\begin{eqnarray}\label{mnp}
m_n^*-m_p^* \approx \frac{2m_s^*}{m_v^*}(m_s^*-m_v^*)\delta.
\end{eqnarray}
Finally, the mean-field potential of a nucleon with infinitely large
momentum in symmetric nuclear matter at the saturation density can
be expressed as
\begin{eqnarray}\label{U0}
U_0^\infty=\frac{A_{l}+A_{u}}{2}+B.
\end{eqnarray}
The values of the parameters $A_0$, $B$, $C_{u0}$, $C_{l0}$, $\sigma$, $\Lambda$, $x$, and $y$ can be obtained from Eqs.~(\ref{E0}-\ref{U0}), with given $\rho_0$, $E_0(\rho_0)$, $K_0$, $U_0^\infty$, $m_s^*$, $E_{sym}(\rho_0)$, $L$, and $m_v^*$.

\end{document}